\documentclass[aps,pra,twocolumn,showpacs,reprint,superscriptaddress,floatfix]{revtex4-1}
\usepackage{amsmath,amssymb,amsfonts}
\usepackage{graphicx}
\usepackage{epstopdf}
\usepackage{mathrsfs}
\usepackage{bm}
\bmdefine\bmzeta{\zeta}
\begin{document}

\title{Magnetic and nematic phases in a Weyl type spin-orbit-coupled spin-1 Bose gas}
\author{Guanjun Chen}
\affiliation{Institute of Theoretical Physics, Shanxi University, Taiyuan 030006, China}
\affiliation{Department of Physics, Taiyuan Normal University, Taiyuan 030001, China}
\author{Li Chen}
\affiliation{Institute of Theoretical Physics, Shanxi University, Taiyuan 030006, China}
\author{Yunbo Zhang}
\email{ybzhang@sxu.edu.cn}
\affiliation{Institute of Theoretical Physics, Shanxi University, Taiyuan 030006, China}
\date{\today}

\begin{abstract}
We present a variational study of the spin-1 Bose gases in a harmonic trap with three-dimensional spin-orbit 
coupling of Weyl type. For weak spin-orbit coupling, we treat the single-particle ground states as the form of 
perturbational harmonic oscillator states in the lowest total angular momentum manifold with $j=1, m_j=1,0,-1$. 
When the two-body interaction is considered, we set the trail order parameter as the superposition of three 
degenerate single-particle ground-states and the weight coefficients are determined by minimizing the energy
functional. Two ground state phases, namely the magnetic and the nematic phases, are identified depending 
on the spin-independent and the spin-dependent interactions. Unlike the non-spin-orbit-coupled spin-1 Bose-Einstein condensate for 
which the phase boundary between the magnetic and the nematic phase lies exactly at zero spin-dependent 
interaction, the boundary is modified by the spin-orbit-coupling. We find the magnetic phase is featured with 
phase-separated density distributions, 3D skyrmion-like spin textures and competing magnetic and biaxial nematic
orders, while the nematic phase is featured with miscible density distributions, zero magnetization and spatially
modulated uniaxial nematic order. The emergence of higher spin order creates new opportunities for exploring
spin-tensor-related physics in spin-orbit coupled superfluid.

\end{abstract}


\maketitle

\section{Introduction}
With the two-photon Raman coupling technique, a synthetic spin-orbit (SO) coupling has been realized
in the pseudo-spin-1/2 Bose-Einstein condensates (BECs) \cite{lin2011,Galitski2013}. Since then, many
theoretical \cite{ho2011,Li2012,martone2012,Li2013,lu2013,Zheng2013,ozawa2013,martone2014,zhai2015} and
experimental \cite{zhangjy2012,qu2013,ji2014,khamehchi2014,engels2014} works have been focused on this
field including the unveiling of the well-known plane wave, stripe and zero-momentum phases \cite{Li2012}. 
In addition, the Raman-induced SO coupled spin-1 BEC has also been realized recently \cite{spielman2015} 
which is unattainable in condensed matter materials, and spatially modulated nematic order is expected to 
appear in the stripe phase \cite{lan2014,natu2015,sun2015,martone2015} .

The Raman-induced SO coupling is a one-dimensional (1D) configuration as an equally weighted Rashba and 
Dresselhaus couplings. This year also witnessed the experimental progress in engineering the two-dimensional (2D) 
Rashba-Dresselhaus-type SO coupling in cold atom gases \cite{zhang2015,pan2015}. 
Many interesting properties have been predicted for the Rashba SO coupled Bose gases
\cite{wang2010,wu2011,sinha2011,zhai2012,hu2012prl,Ramachandhran2012,yu2013,zhou2013,Ramachandhran2013,
ozawa2012,ozawa2012-2,zhangyp2012,cui2013,xu2013,nikolic2014,hu2012prl,Juzeliunas2010,hu2012,wen2012,
song2014,chen2014}, among which the weakly coupled BEC is found to condense into various half-quantum vortex 
phases \cite{wu2011,sinha2011,hu2012prl,Ramachandhran2012,chen2014} due to modification of the single particle spectrum 
by the trapping potential. All these efforts make the cold atom gases with the synthetic gauge field a rapidly 
developing field.

Researchers also go further to deal with the three-dimensional (3D) analogy of
Rashba configuration, i.e. the Weyl SO coupling \cite{wu2012,shenoy2012,Kawakami2012,Cui2012,wu2013,
zhang2013,clark2013,gupta2013,chen2015,liao2015} and the experimental schemes for engineering the Weyl 
configuration in the ultracold gases have been proposed \cite{Liy2012,Anderson2012,Anderson2013}.
While the Weyl coupled spin-1/2 bosons in a homogeneous system reproduces the plane wave and the 
stripe phases \cite{liao2015}, the 3D skyrmion mode with magnetic order
\cite{wu2012,Kawakami2012,wu2013,chen2015} spontaneously appears in the ground state of a trapped system.

For a trapped Weyl coupled spin-1 BEC, one expects the emergence of topological objects with even higher spin 
order, i.e. the nematic order. An immediate question is that, what will the phase diagram look like, and in what a 
way will the 3D skyrmion and (or) the nematic order manifest themselves in the individual phases. Here 
we consider a spin-1 bosonic system subject to a weak Weyl SO coupling of $\mathbf{s}\cdot\mathbf{p}$ type 
in a harmonic trap. We implement a standard variational approach \cite{ho2011,Li2012,wang2010,wu2011,chen2015} 
to give a phase diagram of the system in different interaction regimes. A magnetic phase and a nematic phase are
predicted in the ground state, and the latter is entirely new and
has no analogue in the 3D SO coupled pseudo-spin-1/2 bosonic gases \cite{wu2012,Kawakami2012,wu2013,chen2015}.

The paper is organized as follows. In Sec. II the energy functional for the 3D SO coupled spin-1 condensate is
introduced in harmonic oscillator units. In the weak coupling limit we construct our variational order parameter
in Sec. III and calculate the energy functional by means of the irreducible tensor method.
The ground state phase diagram is determined by numerically minimizing the energy functional
with respect to the variational parameters, and Sec. IV is devoted to an explicit illustration and detailed discussion
of the densities and spin orders in the two ground state phases. We summarize our main results in Sec. V.

\section{\label{model}Model}
We start from the mean-field Gross-Pitaevskii (GP) energy functional of spin-1 bosons with
a Weyl type 3D SO coupling in the presence of a harmonic trap
\begin{equation}
\mathcal {E}=\mathcal {E}_{0}+\mathcal {E}_{int},\label{gpfunctional}
\end{equation}
where the single particle part is%
\begin{equation}
\mathcal {E}_{0}=\int d^{3}\mathbf{r}\Psi^{\dag}\left(
\mathbf{r}\right)
\left(  \frac{\mathbf{p}^{2}}{2m}+\frac{1}{2}m\omega^{2}r^{2}%
+\lambda\mathbf{s}\cdot\mathbf{p}\right) \Psi\left( \mathbf{r}\right),\label{e0}
\end{equation}
with $m$ the atomic mass and $\omega$ the trap frequency. $\Psi=(\psi_{1},\psi_{0},\psi_{-1})^{T}$
denotes the spinor order parameters for bosons with hyperfine components $1,0,-1$ respectively,
and $\mathbf{s}=(s_x,s_y,s_z)$ are spin-1 matrices.
The Weyl SO coupling $\mathbf{s}\cdot\mathbf{p}$ with strength $\lambda$ is a 3D
analogy of the Rashba configuration \cite{Anderson2012}. Without the trapping potential, the momentum $\mathbf{p}$ and the helicity $\mathbf{s\cdot p/}\left\vert \mathbf{p}\right\vert $ are constants of motion, and the single-particle ground states are highly degenerate, i.e. the lower-energy
helical branch achieves a minimum along a sphere of radius $p_{SO}=m\lambda$, called SO sphere \cite{wu2012}.
The presence of a trapping potential will \textit{partially} lift this
degeneracy, but at least a two-fold degeneracy related to the time-reversal
symmetry will still remain \cite{Galitski2013}. The single particle Hamitonian is also invariant
under simultaneous rotation of spin and coordinate space $SO(3)_{R+S}$ that leaves
the total (instead of spin) angular momentum a good quantum number \cite{Kawakami2012}. This breaking of
rotational symmetry in spin space leads to spin-textured ground states
with magnetic and nematic orders, as previously studied in Ref. \cite{wang2010,martone2015,wu2011,natu2015,chen2015}.

The interaction energy functional is formulated
in the standard form \cite{ho1998,ohmi1998,law1998}
\begin{equation}
\mathcal {E}_{int}=\frac{1}{2}\int d^{3}\mathbf{r}\left(
c_{0}n^{2}+
c_{2}\mathbf{S}^{2}\right),
\end{equation}
where $n\left(  \mathbf{r}\right)  =n_{1}\left(  \mathbf{r}\right)
+n_{0}\left(  \mathbf{r}\right)  +n_{-1}\left(  \mathbf{r}\right)  $
is the total particle density and $n_{1,0,-1}\left(  \mathbf{r}\right)
=\left\vert \psi_{1,0,-1}\left(  \mathbf{r}\right)  \right\vert
^{2}$ are densities for the three components, respectively, and
$\mathbf{S}=\Psi^{\dag }\mathbf{s}\Psi$
is spin density. $c_{0,2}$ are spin-independent and spin-dependent interaction strengthes respectively,
which are related to the two-body scattering lengths in the total spin-0 and spin-2 channels as
$c_0=4\pi(a_0 +2a_2)/3m$ and $c_2=4\pi(a_2 -a_0)/3m$.
The interaction is time-reversal (TR) symmetric under
the operation $T=e^{-i\pi s_{y}}K$ with $K$
the complex conjugate. Besides, this
two-body interaction is also SU(2) spin-rotation symmetric, which is different
from the spin-$1/2$ bosons \cite{wu2012,Kawakami2012,wu2013,chen2015}. In the latter case the two body interaction is SU(2)
symmetric only under the condition $g_{\uparrow \uparrow }=g_{\downarrow
\downarrow }=g_{\uparrow \downarrow }$, i.e. $c=g_{\uparrow \downarrow
}/g_{\uparrow \uparrow }=1$. Later we will see that this highly symmetric
interaction will lead to highly degenerate ground states.
In harmonic oscillator units,
the system has length scale $l_{T}=\sqrt{\hbar/m\omega}$, energy
scale $\hbar\omega$, 
and SO coupling strength is in unit of
$\sqrt{\hbar\omega/m}$. If we normalize the order parameter to unity, i.e.,
$\Psi \rightarrow \sqrt{N/l_{T}^{3}}\Psi$ with $N$ the total particle number in the
condensate, the energy functional per particle is obtained as
\begin{align}
\varepsilon &  =\int d^{3}\mathbf{r}\Psi^{\dag}\left(  \mathbf{r}%
\right)  \left\{
-\frac{\nabla^{2}}{2}+\frac{r^{2}}{2}+\lambda\mathbf{s}\cdot\mathbf{p}
\right\}  \Psi\left(  \mathbf{r}\right) \nonumber\\
&  +\int d^{3}\mathbf{r}\left(
\frac{c_{0}}{2}n^{2}+\frac
{c_{2}}{2}\mathbf{S}^{2}\right). \label{Efunctional}%
\end{align}

\section{Variational Approach}
Here we first introduce the single-particle ground states in the form of perturbational
3D harmonic oscillator in the weak SO coupling limit, the superposition of which
constitutes the variational order parameter. We then calculate
the energy functional of Eq. (\ref{Efunctional}) using the irreducible tensor method.

\subsection{Variational Order Parameter}

The variational order parameter is just
a spin-1 version of that presented in our recent work on the 3D SO coupled pseudo-spin-1/2 system \cite{chen2015}. We include some necessary contents here for completeness.
It is expected that in the case of weak SO coupling the single particle energy is dominated by the 3D harmonic oscillator part, which has been verified
in Ref. \cite{wu2012,clark2013,chen2015}.
The solution of the 3D harmonic oscillator is well-known
with the energy eigenvalues $\epsilon_{n_rl}=2n_{r}+l+\frac{3}{2}$ and
eigenfunctions $\phi_{n_{r}lm_l}\left(  r,\theta,\varphi\right)  =R_{n_{r}%
l}\left(  r\right)  Y_{lm_l}\left(  \theta,\varphi\right)$. Here $R_{n_{r}l}$ is the radial wave function and $Y_{lm_l}$ is the
spherical harmonics with
$n_{r} $ the radial quantum number, $l$ the orbital angular
momentum quantum number, and $m_l$ its magnetic quantum number.
In order to take into account
the SO coupling term $\mathbf{s}\cdot\mathbf{p}$ later, it is
convenient to choose the coupled representation of angular momentum for spin-1 particles,
i.e. the complete set of commutative operators includes $\mathbf{l}^2,\mathbf{s}^2,\mathbf{j}^2,j_{z}$
where $\mathbf{j}=\mathbf{l}+\mathbf{s}$ and $j_{z}$ denote the total angular
momentum and its $z$-component, respectively. The eigenfunction
$\phi_{n_{r}lm_l}\left( r,\theta,\varphi\right) $ should be combined with
the spin wave function $\chi_{m_s} (m_s =1,0,-1)$ in the coupled representation as%
\begin{equation}
\phi_{n_{r}ljm_{j}}\left(  r,\theta,\varphi\right) =R_{n_{r}l}\left(
r\right) Y_{jm_{j}}^{l}\left( \Omega\right),
\end{equation}
where $Y_{jm_{j}}^{l}\left( \Omega\right)  =\sum_{m_{l},m_{s}}
{\cal C}_{lm_l1m_s}^{jm_j} Y_{lm_l} \chi_{m_s}$ is the
vector spherical harmonics \cite{varshalovich1988quantum} with
$j=l+1,l,|l-1|$ (if $l=0$, $j=1$ only) and ${\cal C}_{lm_l1m_s}^{jm_j} $ the Clebsch-Gordan coefficients.
In the coupled representation, the ground state wave function has $n_r=l=0$.
This gives a total angular momentum $j=1$ with
$m_{j}=1,0,-1$ and the three ground states are
\begin{equation}
\phi_{001\pm1}\left(
\mathbf{r} \right) =R_{00}\left(  r\right)  Y_{1\pm1%
}^{0}\left(  \Omega\right)
\end{equation}
and
\begin{equation}
\phi_{0010}\left(
\mathbf{r} \right) =R_{00}\left(  r\right)  Y_{10%
}^{0}\left(  \Omega\right)
\end{equation}
respectively.
 \begin{figure}[t]
 \centering{
 \includegraphics[width=0.7\columnwidth]{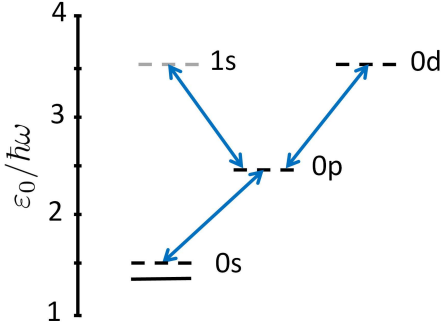}
 }
 \caption{\label{single}(Color online). The lowest energy levels of 3D harmonic oscillator (dashed horizontal lines) and the SO coupled single-particle ground states (solid horizontal line). The selection rules for transition between states with different parities are indicated by the two-way blue arrows. The excited states such as $1s$ (gray dashed line) is neglected in the calculation.}
\end{figure}
The lowest few levels of the 3D harmonic oscillator is shown in Fig. \ref{single}.
The single particle spectrum may be understood as a weak perturbation (slight level
mixing) of the harmonic-oscillator energy levels in the case of
weak SO coupling, which conserves the total angular momentum $\mathbf{j}$ and
$j_{z}$ but $l$ is no longer a good
quantum number. This means SO coupling term would couple $s$, $p$ and even $d$ waves
into the ground state with the same $\mathbf{j}$ and $j_{z}$ as illustrated in Fig. \ref{single}.
In the case of strong SO coupling, the energy spectrum is weakly dispersive or nearly flat
\cite{Ramachandhran2013,wu2012,clark2013} which will not be considered in this work.
The ground state wave functions are thus the superposition of the lowest $s$, $p$ and $d$ wave
states with $j=1$. The state with $m_{j}=1$ is
\begin{equation}
\Phi _{j=1,m_{j}=1}=A_{0}\phi _{0011}+iA_{1}\phi _{0111}-A_{2}\phi _{0211}.  \label{splusp}
\end{equation}
Here $A_{l}\left( l=0,1,2\right) $ are the weight coefficients with the
normalization constraint $\sum_{l=0}^{2}\left\vert A_{l}\right\vert ^{2}=1$, and $i$ in front of $A_{1}$
comes from the pure imaginary matrix elements of SO coupling between $\phi _{0011}$ ($\phi _{0211}$) and $\phi _{0111}$ \cite{wu2012,clark2013,chen2015}. Explicitly this state is a spinor
\begin{align}
&\Phi_{j=1,m_{j}=1}\nonumber \\
&= \left(
\begin{array}{c}
A_{0}R_{00}Y_{00}-iA_{1}\sqrt{\frac{1}{2}}R_{01}Y_{10}-A_{2}\sqrt{\frac{1}{10}}R_{02}Y_{20} \\
iA_{1}\sqrt{\frac{1}{2}}R_{01}Y_{11}+A_{2}\sqrt{\frac{3}{10}}R_{02}Y_{21} \\
-A_{2}\sqrt{\frac{3}{5}}R_{02}Y_{22}
\end{array}
\right).
\end{align}
Moreover, in the single particle level, the energies are irrelevant
to the magnetic quantum number of $j_{z}$. Thus the other two states with $m_{j}=0$ and $-1$
are
\begin{align}
& \Phi _{j=1,m_{j}=0}\nonumber \\
& =\left(
\begin{array}{c}
-iA_{1}\sqrt{\frac{1}{2}}R_{01}Y_{1-1}-A_{2}\sqrt{\frac{3}{10}}R_{02}Y_{2-1}
\\
A_{0}R_{00}Y_{00}+A_{2}\sqrt{\frac{2}{5}}R_{02}Y_{20} \\
iA_{1}\sqrt{\frac{1}{2}}R_{01}Y_{11}-A_{2}\sqrt{\frac{3}{10}}R_{02}Y_{21}%
\end{array}%
\right)
\end{align}%
and
\begin{align}
& \Phi _{j=1,m_{j}=-1} \nonumber \\
& =\left(
\begin{array}{c}
-A_{2}\sqrt{\frac{3}{5}}R_{02}Y_{2-2} \\
-iA_{1}\sqrt{\frac{1}{2}}R_{01}Y_{1-1}+A_{2}\sqrt{\frac{3}{10}}R_{02}Y_{2-1}
\\
A_{0}R_{00}Y_{00}+iA_{1}\sqrt{\frac{1}{2}}R_{01}Y_{10}-A_{2}\sqrt{\frac{1}{10%
}}R_{02}Y_{20}%
\end{array}%
\right).
\end{align}%
respectively. Note that the state $\Phi_{j=1,m_{j}=-1}$ is the time reversal of
$\Phi_{j=1,m_{j}=1}$, and $T\Phi_{j=1,m_{j}=0}=-\Phi_{j=1,m_{j}=0}$. We have neglected
contribution from the excited $1s$ wave state $\phi _{1011}$ shown in Fig. \ref{single}
(gray dashed), which can be absorbed into the lowest $0s$ state $\phi
_{0011}$ owing to the same angular-spin wave function. It has been verified that
inclusion of this excited $1s$ state in the calculation will not alter our main conclusion
for weak two-body interaction.

In the language of Raman-induced SO coupling \cite{lin2011,Zheng2013,ji2014}, the single particle ground state has three minima corresponding to three states $ \Phi _{j=1,m_{j}=0,\pm 1}$.
At the many-body level, the interaction will determine which minimum or
minima the Bose gases will condensate to by minimizing the GP energy
functional \cite{lan2014}. We therefore set the variational
wavefunction as
\begin{align}
\Psi & =c_{1+}\Phi _{j=1,m_{j}=1}+c_{10}\Phi _{j=1,m_{j}=0}+c_{1-}\Phi
_{j=1,m_{j}=-1}\label{variationop}
\end{align}%
with the normalization constraint $c_{1+}^{2}+c_{10}^{2}+c_{1-}^{2}=1$ that ensures the
conservation of the particle numbers. The coefficients $c_{1+},c_{10},c_{1-}$, to be
determined by the interaction, are generally complex. For the sake of simplicity,
we restrict them to be real here. Later, we will discuss the consequences of
such restriction. This variational wave function ansatz is extensively encountered in
SO coupled cold atom gases. \cite{ho2011,Li2012,wang2010,wu2011,zhai2012,yu2013,lan2014,natu2015,liao2015}

\subsection{Nematic Order}\label{nematic}

Prior to the calculation of the energy functional, we first elucidate the meaning of our variational order parameter ansatz. Since our single-particle hamiltonian respects the SO(3)$_{R+S}$ symmetry, it is convenient to introduce the polarization operator \cite{varshalovich1988quantum} to describe the spin order. The polarization operators for spin-$s$ system $T_{m_{l}}^{\left( l\right) }(s)$ with $l=0,1,\ldots ,2s$ and $m_{l}=-l,-l+1,\ldots l$, are a set of $\left(2s+1\right)^2$ operators which act on the spin wave functions
and transform under the coordinate system rotation according to the irreducible
representation of SO(3) group. In this sense, it is also an
irreducible tensor of rank $l$.

For spin-1 objects, the nine polarization operators $T_{m_{l}}^{\left( l\right) }(s)$ with $l=0,1,2$
constitute a complete set of square $3\times 3$ matrices and are generators of the unitary
group $U(3)$ in rotationally covariant Racah form \cite{iachello2006}. The rank-$0$ operator is the unit $3 \times 3$ matrix $I$
\begin{equation}
T_{0}^{\left( 0\right) }(s)=\frac{1}{\sqrt{2s+1}}I.
\end{equation}%
The rank-$1$ operators $T_{m_{l}}^{\left( 1\right) }(s)$ are proportional
to the irreducible rank-1 spin tensor with components
\begin{equation}
T_{m_{l}}^{\left( 1\right) }(s) =\frac{\sqrt{3}}{\sqrt{s\left( s+1\right)
\left( 2s+1\right) }}s^{(1)}_{m_{l}},\left( m_{l}=0,\pm 1 \right)
\end{equation}
where the spherical components of the irreducible rank-1 spin tensor
$s^{(1)}_{m_{l}}$ are related to the cartesian ones as
\begin{equation}
s^{(1)}_{\pm 1} =\mp \frac{1}{\sqrt{2}}\left( s_{x}\pm is_{y}\right) ,s^{(1)}_{0}=s_{z}.
\end{equation}
The rank-$2$ polarization operators $T_{m_{l}}^{\left(2\right) }(s)$ are
\begin{align}
T_{m_{l}}^{\left( 2\right) }(s)& =\sum_{\mu+\nu=m_l}{\cal C}_{1\mu 1\nu
}^{2m_{l}}s^{(1)}_{\mu }s^{(1)}_{\nu } \nonumber\\
& =\left\{ s^{\left( 1\right) }s^{\left( 1\right) }\right\} _{m_{l}}^{\left(
2\right) },\left( \mu ,\nu =0,\pm 1\right),
\end{align}%
where $\left\{A^{(m)}B^{(n)}\right\}^{(k)}$ defines the rank-$k$ tensor product of rank-$m$ irreducible tensor $A^{(m)}$ and rank-$n$ irreducible tensor $B^{(n)}$.
$T_{m_{l}}^{\left( 2\right) }(s)$
are equivalent to a symmetric \textit{traceless} cartesian tensor of rank-2, i.e. the
spin nematic tensor or quadrupole tensor $\mathcal{N}$ through the relation
\cite{varshalovich1988quantum}
\begin{align*}
T_{\pm 2}^{\left( 2\right) }& =\frac{1}{2}(\mathcal{N}_{xx}-\mathcal{N}_{yy}\pm 2i\mathcal{N}_{xy}), \\
T_{\pm 1}^{\left( 2\right) }& =\mp (\mathcal{N}_{xz}\pm i\mathcal{N}_{yz}), \\
T_{0}^{\left( 2\right) }& =\sqrt{\frac{3}{2}}\mathcal{N}_{zz},
\end{align*}%
where
\begin{align}
\mathcal{N}_{ij}& =\frac{1}{2}\left( s_{i}s_{j}+s_{j}s_{i}-\frac{4}{3}\delta
_{ij}I\right) ,\left( i,j=x,y,z\right). \label{nematictensor}
\end{align}%
The unit matrix $I$ is a spin-rotation invariant scalar which contains important information regarding the {\it charge} (density) order, three matrices $s_{x},s_{y},s_{z}$ form a vector which represents the local spin ({\it magnetic}) order, and five matrices $\mathcal{N}_{ij}$ form a symmetric traceless tensor which represents the local spin
fluctuations or {\it nematic} order \cite{mueller2004,Kawaguchia2012}.
The spin-1 systems
therefore support spin nematic order in addition to the charge and magnetic ones.
The magnetic and the nematic orders compete with each other as increasing one of them requires reducing the other \cite{mueller2004}.

With the polarization operators, our variational wavefunction Eq. (\ref{variationop}) can be written as%
\begin{equation}
\Psi =U\left(\mathbf{r}\right) \bmzeta.\label{variationop3}
\end{equation}
where $\bmzeta=\left( c_{1+},c_{10},c_{1-}\right)^{T}$ is a normalized spinor and the position-dependent transformation
\begin{equation}
U(\mathbf{r})=\sqrt{\frac{2s+1}{4\pi }}\sum_{l=0}^{2s}\left( -i\right)
^{l}A_{l}R_{0l}C^{\left( l\right) }\cdot T^{\left( l\right) } \label{variationop2}
\end{equation}
on the spinor $\bmzeta$ leads to the spin-textured ground states in the SO coupled cold atom
gases. Here the dot defines the scalar product of the modified spherical harmonics $C^{(l)}$
and $T^{(l)}$. This local modulation operator is expanded in series of $C^{\left( l\right) }\cdot
T^{\left( l\right) }$ with the highest order $l=2s$. The
hamiltonian is invariant under the simultaneous rotation in spin and coordinate
space $SO(3)_{R+S}$, therefore trapped spinor condensate with SO coupling inevitably
carries angular momentum by twisting its spinor order parameter \cite{mueller2004,chen2015}.

It is also intuitive to understand the modulation of spin-1/2 objects discussed in detail in Ref.
\cite{chen2015}, in which case the four polarization operators $T_{m_{l}}^{\left( l\right) }(s)$
with $l=0,1$, or explicitly $I,s^{(1)}_{+1},s^{(1)}_{0},s^{(1)}_{-1}$, constitute a complete set of square
$2\times 2$ matrices. The transformation matrix $U(\mathbf{r})$ can be rewritten as
\begin{equation}
U(\mathbf{r})=\sqrt{\tilde{n}(r)}e^{-i\omega(r)\hat{\mathbf{r}}\cdot \mathbf{s}}\label{spinhalfu}
\end{equation}
where the density and spin modulations are apparently separated owing to the absence of
nematic order. The spin-1 system thus enables to explore spin-tensor-related physics in the SO
coupling superfluid, which has fundamentally different rotation properties as in spin-1/2 system.

\subsection{Energy Functional}\label{functional}

We now have six variational parameters $A_0, A_1, A_2$, $c_{1+}, c_{10}, c_{1-}$ in the trail
variational order parameter. The energy functional of Eq. (\ref{Efunctional}) can be calculated
analytically using the proposed order parameter Eq. (\ref{variationop}).

The single particle part of the energy functional consists of the kinetic energy, the trapping
potential of the 3D harmonic oscillator, and the SO coupling term. As in the spin-1/2 case
\cite{chen2015}, the matrix elements for the kinetic energy and trapping potential are
non-vanishing for states with the same
parity, while the SO coupling term $\mathbf{s}\cdot \mathbf{p}$
will mix states with opposite parities. The result is
\begin{align}
& \int d^{3}\mathbf{r}\Psi^{\dag}\left(  \mathbf{r}%
\right)  \left\{
-\frac{\nabla^{2}}{2}+\frac{r^{2}}{2}+\lambda\mathbf{s}\cdot\mathbf{p}
\right\}  \Psi\left(  \mathbf{r}\right) \nonumber\\
&  =\frac{3}{2}A_{0}^{2}+\frac{5}{2}A_{1}^{2}+2\lambda
A_{0}A_{1}+\Delta_0. \label{singlepart}
\end{align}
The contribution from the $d$-wave states is collected in $\Delta_0$ (see Supplemental Material),
and we have used%
\begin{align*}
&\left\langle \phi_{0011}\left\vert
\mathbf{s}\cdot\mathbf{p}\right\vert \phi_{0111}\right\rangle =-i,\\
& \left\langle \phi _{0111}\left\vert \mathbf{s\cdot p}\right\vert \phi
_{0211}\right\rangle =i\sqrt{\frac{5}{6}}.
\end{align*}%
We refer to Ref. \cite{chen2015} for details of the integral calculation
in which the method of irreducible tensor algebra is employed \cite{varshalovich1988quantum}.

The calculation of the interaction is tedious but straightforward which yields
\begin{equation}
\int d^{3}\mathbf{r}\frac{c_{0}}{2}n^{2}
 =\frac{c_{0}}{8\pi}\sqrt{\frac{2}{\pi }}\left[
A_{0}^{4}+A_{0}^{2}A_{1}^{2}+\frac{1}{16}\left( 7+x\right) A_{1}^{4}+\Delta_n\right],
\label{spinindepend0d}
\end{equation}
and
\begin{equation}
\int d^{3}\mathbf{r}\frac{c_{2}}{2}\mathbf{S}^{2}
=\frac{c_{2}}{8\pi}\sqrt{\frac{2}{\pi }}\left[
A_{0}^{4}+A_{0}^{2}A_{1}^{2}+\frac{5}{16}A_{1}^{4}+\Delta_s\right] \left( 1-x\right). \label{spindependent0d}
\end{equation}
with $x=\left[ 1-\left( c_{1+}+c_{1-}\right) ^{2}\right] ^{2}$.
Here $\Delta_n$ and $\Delta_s$ again denote the contributions from the $d$-wave states
(see Supplemental Material). The energy functional per particle is simply the summation of
Eqs. (\ref{singlepart}), (\ref{spinindepend0d}) and (\ref{spindependent0d}).

\section{Ground state phase diagram}

\begin{figure}[t]
\includegraphics[width=0.8\columnwidth]{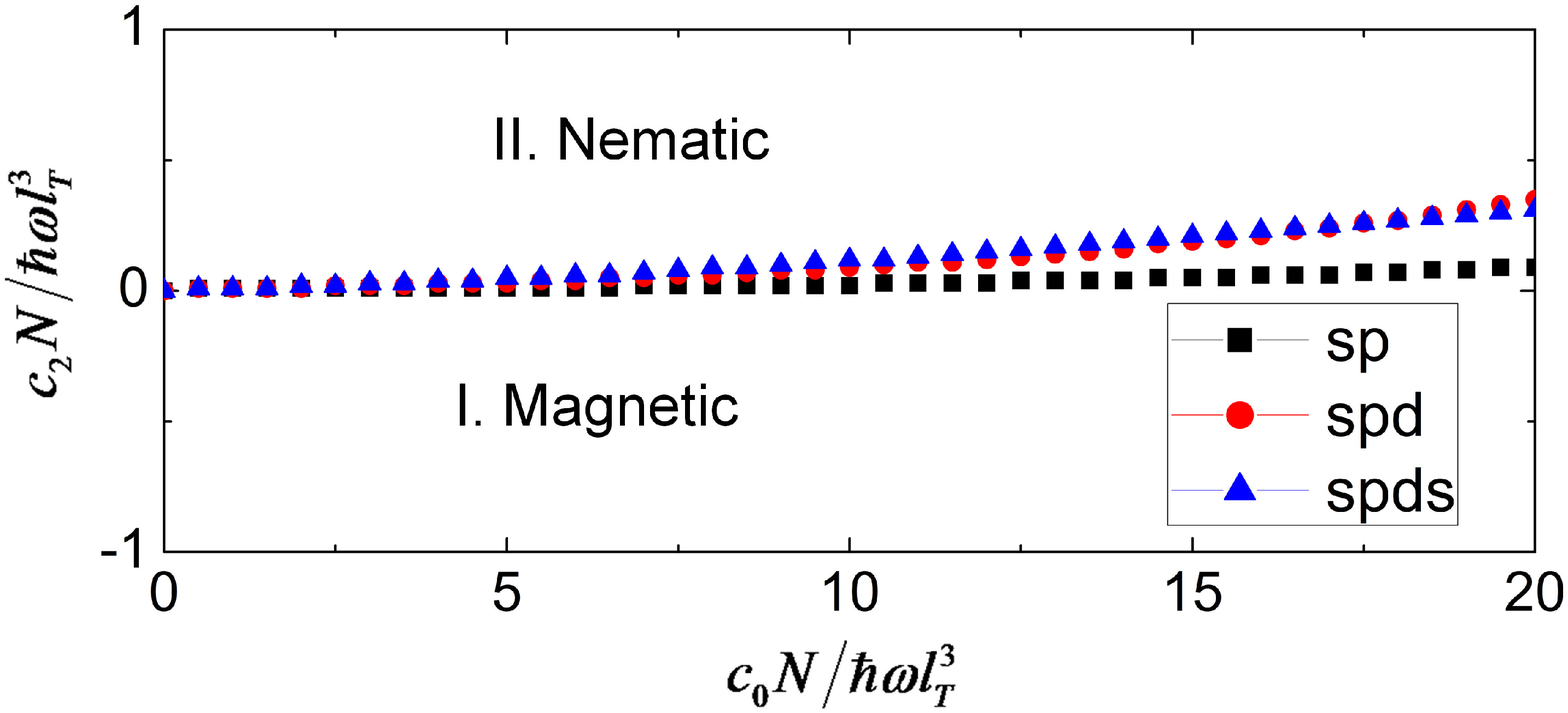}
\caption{Phase diagram of weakly SO coupled spin-1 bosons with coupling strength $\lambda=0.4$.
The boundary between the magnetic phase and the nematic phase is determined in successive approximations in which
we include $0s$, $0p$, $0d$ and $1s$ orbits step-by-step. Density distributions and spin textures of the two phases are shown in Fig. \ref{phase1z} to Fig.  \ref{phase2} respectively.}
\label{phases}
 \end{figure}

The ground state phase diagram for a given coupling strength $\lambda $ is obtained by
minimizing the variational energy with respect to the parameters $%
A_{l}$ and $x$ under two constraints $\sum_{l=0}^{2}\left\vert A_{l}\right\vert ^{2}=1$
and $c_{1+}^{2}+c_{10}^{2}+c_{1-}^{2}=1$. The latter further restricts $x$
to the region $[0,1]$. Should the spin-independent interaction vanish $c_{0}=0$, we
see from Eq. (\ref{spindependent0d}) that the optimized parameter $x$ either takes value of $0$
for $c_{2}<0$, or $1$ for $c_{2}>0$ for negligibly small contribution $\Delta_s$ from the $d$-wave states.
The analysis below will show that this assumption
holds generally except for extremely strong interaction $c_{0}$. The variational ansatz characterizes
two quantum phases: (I) the magnetic phase with a ferromagnetic manifold $\bmzeta$ for $c_{2}<0$,
as $x=0$ means that $\left( c_{1+}+c_{1-}\right) ^{2}=1$; (II) the polar or nematic phase with a polar manifold
for $c_{2}>0$, as $x=1$ means $\left( c_{1+}+c_{1-}\right) ^{2}=0$ or $2$.
This is consistent with the conventional spin-1 BEC \cite{ho1998,ohmi1998,law1998} where for
$c_{2}<0$ ($c_{2}>0$) a ferromagnetic (polar) spinor is needed to minimize the mean-field energies.
If $c_{0}>0$, the boundary between magnetic and nematic phases will drift a little to the positive $c_{2}$ side
because the spin-independent interaction (\ref{spinindepend0d}) which prefers
$x=0$, prevails in the $c_2>0$ regime over the spin-dependent one (\ref{spindependent0d}) which prefers $x=1$.
For the values of optimized parameters $A_{l}$, generally, the weights of $p$ and $d$ waves becomes
more and more important with increasing interaction $c_{0}$, which effectively diminishes the partition
of $s$ wave. In the range of $c_{0}=0\sim 20$ we considered here, $|A_{0}|^2$ decreases
from $0.884$ to $0.723$, $|A_{1}|^2$ increases from $0.110$ to $0.260$, while $|A_{2}|^2$ from $%
0.003$ to $0.014$ which is always negligibly small. A typical phase diagram is shown
in Fig. \ref{phases} where the phase boundary is calculated in three successive approximations: ``sp'' -
only the lowest $0s$ and $0p$ states are considered in the approximation; ``spd'' - the $0d$ energy level
is added; ``spds'' - the $1s$ energy level is added further. We can find that the boundary does not alter
significantly when we include the excited $s$ wave states $\phi _{101m_j}$ in the variational order parameter.

It has been pointed out that the SO coupling manifests itself in a way that the modulation of the ferromagnetic and
polar spin textures in the pseduspin space could be transferred to patterned structures in orbit
space even in the ground states \cite{lan2014}. The reason of this modulation
lies in that, in the presence of SO coupling \cite{xu2011} or dipolar interaction \cite{wilson2013}, the
spin-dependent interaction would inevitably influence the spatial motion, which leads to
rich density pattern. We discuss this in the following for the magnetic and polar phases explicitly.

\subsection{Magnetic Phase}
 \begin{figure}[!t]
 \includegraphics[width=1\columnwidth]{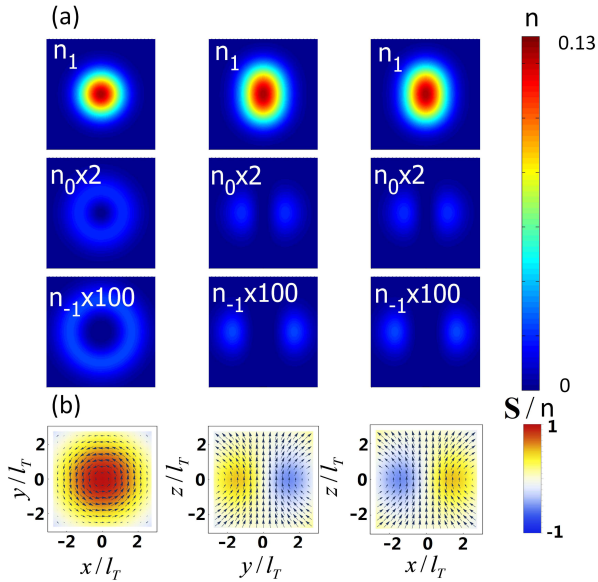}
 \caption{\label{phase1z}(Color online). Density distributions and spin textures of the longitudinally  magnetized state
 $\bmzeta_{z}$. The variational parameters used are $(A_0,A_1,A_2)=(-0.85,0.51,0.12)$. (a) Density distributions. Three columns are densities in $xy$, $yz$ and $xz$ planes respectively. Three rows are density distributions of $+1,0,-1$ components
respectively as explicitly labeled; (b) Spin texture
$\mathbf{S}(\mathbf{r})/n(\mathbf{r})$. The figures plot the spin in $xy$, $yz$, and $xz$ planes respectively. The arrows indicate the in-plane components of the local spin, and the color scale shows the magnitude of the out-plane component. }
 \end{figure}

This phase lies in the lower part of the parameter space with the states featured as $c_{1+}+c_{1-}=\pm 1$.
The corresponding spinor $\bmzeta$ in Eq. (\ref{variationop3}) denotes a ferromagnetic state with magnetization along
any axis in the $xz$ plane. This asymmetry between $x$, $z$ and $y$ directions is a consequence
of our simplified treatment which restricts the coefficients $%
c_{1+},c_{10},c_{1-}$ to be real, such that the spinor $\bmzeta$ is unable
to describe the state with magnetization along $y$ direction. If we
relax this restriction to allow complex coefficients $c_{1+},c_{10},c_{1-}$,
all states with spinors $\bmzeta$ magnetized along any spatial direction
belong to this magnetic phase, which leads to an infinitely degenerate ground states.
Two typical spinors are $\bmzeta_{z}=\left( 1,0,0\right) ^{T}$ and
$\bmzeta_{x}=\left( \tfrac{1}{2},\tfrac{1}{\sqrt{2}},\tfrac{1}{2}\right) ^{T}
$ which are longitudinally and transversely magnetized ferromagnetic states respectively. This
phase spontaneously breaks the time-reversal symmetry which leads to spontaneous
magnetization along the $xz$ plane. The situation is just like spin half
case we have studied before \cite{chen2015}. The difference is that the $x$ and $z$
ferromagnetic states are not degenerate for the spin-half case as the
interaction is not SU(2) symmetric, leaving only two-fold Kramers degeneracy
originating from the time-reversal symmetry there. For spin-1 gases here,
the ground states are infinitely degenerate resulted from the SU(2) symmetric
interaction.

We first consider the longitudinally  magnetized state $\bmzeta_{z}=\left( 1,0,0\right) ^{T}$
with the time reversal state being $\bmzeta_{-z}=\left( 0,0,1\right) ^{T}$. We plot the ground state density
distribution for the three components in Fig. \ref{phase1z} (a), which show clearly cylindrical symmetry.
While the spin component 1 is dominantly occupied in the center, which allows the condensate
to develop a longitudinal magnetization, the components $0$ and $-1$ form two toruses surrounding the central
part. It can be seen that the outermost shell of spin $-1$ density is negligibly small which is entirely
attributed to the involvement of the $d$ wave fraction in the order parameter. To visualize the $d$ wave nature
we need to zoom out 100 times in the density plot. Generally only less populated spin component can
develop $d$-wave characters in the ground states because these complex structure in the
high density spin components would cost too much kinetic energy \cite{deng2012}.
For its time-reversal degenerate state $\bmzeta_{-z}$ the spin components $1$ and $-1$
are inversely populated.

The spin texture of the longitudinally magnetized state is plotted in Fig. \ref{phase1z} (b) where
we find a synchronous modulation between the particle density and the spin density owing to the
breaking of spin rotation symmetry. In the trap center, the spins are aligned along the $z$ axis due to
the dominant occupation of spin-1 component. The successive population of the $0$ and $-1$ components
forms a local spin texture $\bf{S(r)}/n(\bf{r})$ where the spin density vectors deflect continuously to the $xy$
plane away from the center. This is a magnetic skyrmion-like texture similar to the half quantum vortex configuration
in the Weyl SO coupled psedo-spin-1/2 BEC \cite{wu2011,Kawakami2012,chen2015}. The spin vector field lines
develop into a bundle of fountain-like streamlines close to the $z$ axis around which a torus is formed near the
$xy$ plane \cite{chen2015}.

\begin{figure}[!t]
\centering{
\includegraphics[width=0.90\columnwidth]{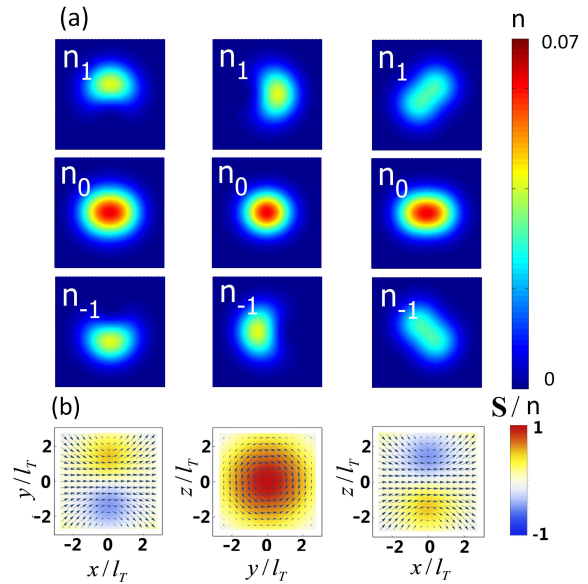}}
\caption{\label{phase1x}(Color online).
Density distributions and spin textures of the transversely magnetized state $\bmzeta_{x}$. The parameters $A$'s are
the same as those of Fig. \ref{phase1z}.}
\end{figure}

The density distribution for the transversely magnetized state $\bmzeta_{x}=\left( \tfrac{1}{2},\tfrac{1}{%
\sqrt{2}},\tfrac{1}{2}\right) ^{T}$ with the time reversal state $\bmzeta_{-x}=\left( \tfrac{1}{2},-\tfrac{1}{%
\sqrt{2}},\tfrac{1}{2}\right) ^{T}$ looks quite differently as shown in Fig. \ref{phase1x} (a).
With half of the atoms filled in the spin-0 component, the components $\pm 1$
becomes equally populated and spatially separated, which leads to a significant transverse magnetization
\cite{deng2012}. The density distributions of three components are
separated in an alternative way, i.e., the $-1$ ($+1$) component lies mainly in the
$-y$ ($+y$) half-space and its peak density center is along the direction joining the III and VIII octants (I and VI octants) \cite{chen2015},
while the $0$ component is embedded between them with a density profile along the $x$ axis.
This leads to a magnetic skyrmion-like texture with the spins in the trap center transversely aligned along the $x$ axis
as shown in Fig. \ref{phase1x} (b), and the increasing population of $\pm1$ components away from the trap center makes the spin
density vector forms a torus near the $yz$ plane. Owing to the non-commutative nature of position-dependent transformation
(\ref{variationop2}) and the spin rotation, the spin texture of the longitudinally magnetized state is different from the $\pi/2$ spin rotation
around $y$ axis of the transversely magnetized state, though the spinor wave functions themselves $\bmzeta_{z}$ and $\bmzeta_{x}$ are
related by such a rotation. All other states degenerate with these two magnetic states have similar properties, except that the
magnetization axes may be in any direction determined by the values of parameters $c$'s.

\subsection{Nematic Phase}
 \begin{figure}[!t]
 \centering{
 \includegraphics[width=0.96\columnwidth]{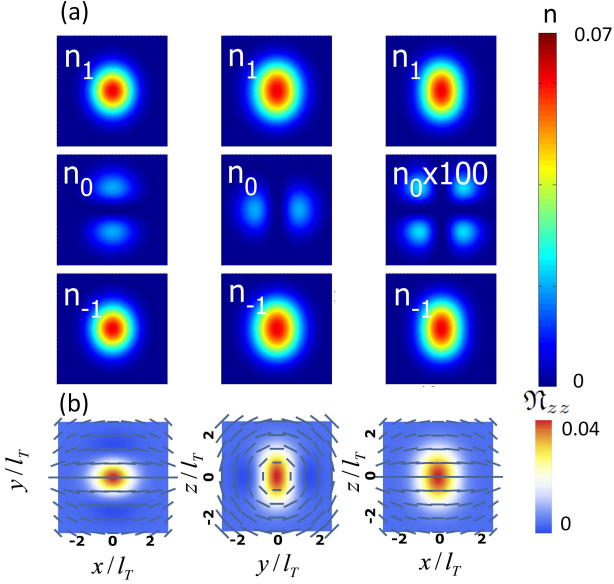}}
 \caption{\label{phase2}(Color online). Density distributions and the nematic directors of the nematic state
 $\bmzeta_{p1}$. (a) Density distributions; (b) Nematic director and the tensor magnetization density $\mathfrak{N}_{zz}$.
 The figures plot the projection of the single nematic director on the $xy$, $yz$, and $xz$ planes respectively.
The parameters $A$'s are the same as those of Fig. \ref{phase1z}.}
 \end{figure}
This phase lies in the upper part of the parameter space with the states featured as $c_{1+}+c_{1-}=0$ or $\sqrt{2}$. The
corresponding spinor $\bmzeta$ in Eq. (\ref{variationop3}) denotes a polar state with zero magnetization $\bf{S(r)}=0$ everywhere.
Two typical spinors are $\bmzeta_{p1}= \frac{1}{\sqrt{2}}\left(1,0,1\right)^{T}$ and $\bmzeta_{p2}= \left(0,1,0\right)^{T}$. The time-reversal symmetry is preserved in this phase, so the ground state has no
spontaneous magnetization as the non-magnetic phase in the Raman induced SO coupled two-component Bose gases
\cite{lin2011,Li2012,Zheng2013,ji2014}. This is a new phase and no analogy in spin half case we considered before
\cite{chen2015}. All states belong to this phase are again infinitely degenerate.

The density distribution of the state $\bmzeta_{p1}$ is plotted in Fig. \ref{phase2} (a). A signature of the $d$-wave is seen
in the 100 times zoomed density of $0$ component in the $xz$ plane, which comes from the $d$-wave contribution in the
variational order parameter. The phase separation among three components, which is possible only in the case of time-reversal
symmetry breaking \cite{gautam2015}, i.e. the magnetic phase, is not seen for this nematic phase which preserves the
time-reversal symmetry and the $\pm 1$ components are miscible.

\begin{figure}[!t]
\includegraphics[width=0.6\columnwidth]{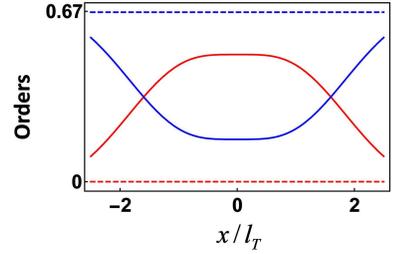}
\caption{\label{competingorders}(Color online). Competing orders in the magnetic and the nematic phases
along $x$ axis. They are $\tfrac{1}{2}\left|\tfrac{\mathbf{S}}{n}\right|^2$(Red) and
$\mathrm{Tr}\left(\frac{\mathfrak{N}}{n}\right)^2$(Blue) respectively. Solid curves are for the magnetic phase
and dashed lines for the nematic phase.} \end{figure}

We define the nematicity density tensor $\mathfrak{N}=\Psi^{\dag}\mathcal{N}\Psi$ to characterize the nematic order
due to the absence of the magnetization. In the nematic phase $\bmzeta_{p1}$ or $\bmzeta_{p2}$,
the normalized nematicity density tensor $\mathfrak{N}/n$ has eigenvalues $\{\tfrac{1}{3},\tfrac{1}{3},-\tfrac{2}{3}\}$
everywhere, therefore describing a uniaxial nematic state.  The eigenvector associated with eigenvalue $-\tfrac{2}{3}$
defines the nematic director which is plotted in Fig. \ref{phase2} (b) together with the tensor magnetization density
$\mathfrak{N}_{zz}=\Psi^{\dag}\mathcal{N}_{zz}\Psi$. The nematic directors form a lantern-like structure with the principle
axis along the $x$ direction. The spatially modulation of the nematic directors, shown as headless
vectors in the $xy$, $yz$ and $xz$ planes, reflects indirectly the modulation of nematicity density tensor themselves. In the
trap center, where the spinor wavefunction $\bmzeta_{p1}$ describes a transverse polar state,
the nematic director points along the $x$ axis. Away from the trap center, the $0$ component is gradually populated
and the nematic director are continuously modulated into concentric circles in the $yz$ plane. For the state
$\bmzeta_{p2}$, the $0$ component in the trap center will allow a longitudinal polar state and
we find an alternative modulation of the nematic directors. The tensor magnetization $\langle\mathfrak{N}_{zz}\rangle$
\cite{spielman2015,sun2015} has been adopted to resolve the order of the phase transition in the Raman-induced SO
coupled spin-1 condensate, and the spatial modulation of $\mathfrak{N}_{zz}$ along direction of the SO coupling
has been noticed \cite{natu2015,martone2015}.

The magnetic phase is also featured with a nematic order and the competing orders in both phases are shown in
Fig. \ref{competingorders}. These two spin orders are competing with each other \cite{mueller2004} to meet the
requirement
\begin{equation}
\frac{1}{2}\left|\frac{\mathbf{S}}{n}\right|^2 + \mathrm{Tr}\left(\frac{\mathfrak{N}}{n}\right)^2 =\frac{2}{3}.
\label{competing}
\end{equation}
This prevents us from writing the transformation
Eq. (\ref{variationop2}) into a local spin rotation as Eq. (\ref{spinhalfu}). Diagonalizing the nematicity density tensor
yields three distinct eigenvalues which are spatially modulated as well as the nematicity density tensor themselves
\cite{martone2015}. Thus the magnetic phase has both magnetic and biaxial nematic orders, while the nematic phase
exhibits only a uniaxial nematic order.

Finally, we remind that our variational order parameter based on
the perturbation expansion may not be applicable in the limit of strong
SO coupling \cite{wu2012,wu2013,hu2012prl,chen2014} where a skyrmion-lattice-like ground state
may appear. Moreover, when two-body interaction is strong enough, the variational order parameter starts to
involve higher angular momentum $j$ states which will break the SO(3)
rotational symmetry, and the Bose gases will condense into the plane wave or stripe phases instead
\cite{sinha2011,wu2012,liao2015}.

\section{Summary}
We establish the ground state phase diagram of the weakly 3D spin-orbit
coupled spin-1 bosons theoretically. The ground state may be in a magnetic or a nematic phase
determined by the competing between the spin-independent and the spin-dependent two-body interaction.
The nematic phase is a new phase that is absent in a 3D spin-orbit coupled pseudo-spin-1/2 bosonic system.
We discuss the density distribution and spin orders of the two phases in detail. The magnetic phase permits both
a magnetic order and a biaxial nematic order, while the nematic phase is featured by a uniaxial nematic order.
These novel phases are in current experimental reach benefiting from the rapid progress of cold gases with artificial
gauge field.

\begin{acknowledgments}
This work is supported by NSF of China under Grant Nos. 11234008 and
11474189, the National Basic Research Program of China (973 Program)
under Grant No. 2011CB921601, Program for Changjiang Scholars and
Innovative Research Team in University (PCSIRT)(No. IRT13076).
\end{acknowledgments}


\end{document}


\title{Supplemental Material: Magnetic and nematic phases in a Weyl type spin-orbit-coupled spin-1 Bose gas}
\author{Guanjun Chen}
\affiliation{Institute of Theoretical Physics, Shanxi University, Taiyuan 030006, China}
\affiliation{Department of Physics, Taiyuan Normal University, Taiyuan 030001, China}
\author{Li Chen}
\affiliation{Institute of Theoretical Physics, Shanxi University, Taiyuan 030006, China}
\author{Yunbo Zhang}
\email{ybzhang@sxu.edu.cn}
\affiliation{Institute of Theoretical Physics, Shanxi University, Taiyuan 030006, China}
\date{\today}
\maketitle

\onecolumngrid

\setcounter{equation}{0}
\setcounter{page}{1}
\makeatletter
\renewcommand{\theequation}{S\arabic{equation}}

\begin{align}
\Delta_0&=\frac{7}{2}A_{2}^{2}-2\lambda A_{1}A_{2}\sqrt{\frac{5}{6}}.\\
\Delta_n&=-\sqrt{\frac{1}{12000}}(10A_{0}A_{1}^{2}A_{2}-7A_{0}A_{2}^{3})\left( 1+3x\right) +\frac{3}{10}A_{0}^{2}A_{2}^{2}\left(2+x\right) +\frac{7}{240}A_{1}^{2}A_{2}^{2}\left( 19-3x\right) +\frac{63}{1600}A_{2}^{4} \left( 7+x\right).\\ 
\Delta_s&=\sqrt{\frac{1}{480}}(10A_{0}A_{1}^{2}A_{2}-7A_{0}A_{2}^{3}) +\frac{35}{48}A_{1}^{2}A_{2}^{2}+ \frac{63}{320}A_{2}^{4}.  
\end{align}